# Dispersion and spin wave "tunneling" in nano-structured magnetostatic spin waveguides.

A. Kozhanov, D. Ouellette, M. Rodwell, S. J. Allen
*California Nanosystems Institute, University of California at Santa Barbara, Santa Barbara, CA, 93106*
A. P. Jacob
*Technology and Manufacturing Group,*
*Intel Corporation, Santa Clara, CA 95052 & Western Institute of Nanoelectronics (WIN), UCLA, Los Angeles, CA 90095*
D. W. Lee and S. X. Wang
*Department of Materials Science and Engineering, Sanford University, Stanford, CA, 94305*

Magnetostatic spin wave dispersion and loss are measured in micron scale spin wave-guides in ferromagnetic, metallic CoTaZr. Results are in good agreement with model calculations of spin wave dispersion. The measured attenuation lengths, of the order of 3 µm, are several of orders of magnitude shorter than that predicted from eddy currents in these thin wires. Spin waves effectively "tunnel" through air gaps, produced by focused ion beam etching, as large as 1.5 µm.

(PACS: 76.50.+g)

Small scale magnetostatic spin wave devices are potentially important for various applications such as on-chip tunable filters for communication systems[1], inductors[2,3] and spin wave logic devices[4,5]. Magnetostatic spin wave devices based on insulating ferrimagnetic materials like yttrium-iron garnet (YIG)[6] have been intensively explored and developed. However, significantly higher saturation magnetization of ferromagnetic metals like CoFe, NiFe and CoTaZr[7] as well as ease of film deposition and processing at the micron and nanometer scale has made these materials potentially more important.

Strong shape anisotropy arises when the ferromagnetic metal films are patterned into wires or stripes. This results in finite frequency spin wave modes with little or no biasing magnetic fields; the dispersion of the backward volume magnetostatic spin waves (BVMSW's[8]) are determined by the profile dimensions. Ferromagnetic wires with magnetization, wave vector k and biasing magnetic field H aligned along the wire have been explored theoretically[9,10] while elegant and powerful Brillouin light scattering experiments document thermally excited lateral standing wave patterns associated with the various spin wave modes.[11,12,13].

Here we use shorted coplanar waveguides[14,15] to excite and detect BVMSW's in spin waveguides and directly measure the spin wave dispersion and attenuation. Further, we document spin wave "tunneling" through gaps in the wires produced by focused ion beam etching.

Recent work has shown "tunneling"[16] and resonant "tunneling"[17] of spin waves in YIG films with controlled magnetic field inhomogeneities. While the analogy to quantum mechanical tunneling of particles and leaking of electromagnetic waves in complex dielectric structures is very useful, it is also recognized that the long range dipolar forces that are an essential feature of magnetostatic spin waves add complexity[16]. The gaps explored here have no magnetic material; there is no tunneling in the conventional sense, only the long range dipolar fields are at play.

$Co_{90}Zr_5Ta_5$ ferromagnetic films, 110 nm thick, were grown by sputter deposition on a Si/SiO$_2$ substrate. A vibrating sample magnetometer measured a saturation magnetization of $M_s \approx 1.2\,\text{T}$ and a coercive field of $H_C \approx 2\,\text{Oe}$ in the unpatterned film. Wires were produced with a Panasonic ICP etch system using chlorine chemistry. An insulating SiO$_2$ layer covered the patterned ferromagnetic wires. Magnetostatic spin waves were excited and detected by coupling loops formed by the short-circuited ends of coplanar waveguides. (Figure 1.) High frequency currents in the signal line of the coplanar waveguide, aligned atop the ferromagnetic wire, produced magnetic fields that excite the magnetostatic spin waves. A focused ion beam system (FIB) was used for making an air gap in the middle of CoTaZr wire perpendicular to the longest dimension (Figure 1). The gap widths varied from 95 nm to 1.5 µm.

S-parameters were measured at room temperature using Agilent 8720ES vector network analyzer. Only $S_{12}$, the ratio of high frequency voltage at terminals 1 to the input high frequency voltage at terminals 2, is analyzed in the following discussion. The test devices were positioned on the narrow gap of small electro-magnet that provided bias up to 1000 Oe. By comparing the S-parameters at disparate bias magnetic fields, the magnetic field independent instrument response can be effectively

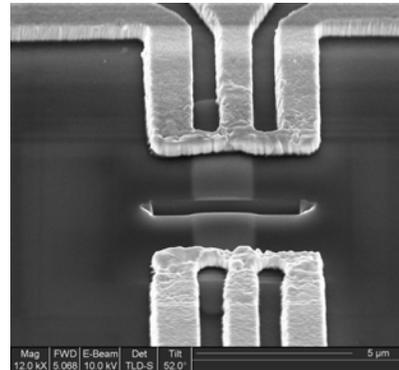

FIG.1. SEM image of a (10×1.8×0.11) µm³ CoTaZr spin wave-guide excited and detected by shorted coplanar waveguides with a 1.2 µm dielectric (air) gap in the center.

removed to expose the S - parameters of the magneto-static spin wave guide.

Typical results for $S_{12}$ measured on a $(10\times1.8\times0.11)$ μm$^3$ CoTaZr stripe at finite magnetic field are shown in Figure 2. At least two reproducible series of oscillations is seen on the real and imaginary parts spectrum of $S_{12}$ (Figure 2a). These features shift towards higher frequencies when the external magnetic field is increased. $Re(S_{12})$ and $Im(S_{12})$ oscillations are $\sim \pi/2$ radians out of phase; they are used to measure the continuous evolution of phase of $S_{12}$ with frequency.

The magnitude and phase of $S_{12}$ are shown in Figure 2b. The two strong features represent BVMSV modes with different spatial distribution of magnetic excitation across the CoTaZr wire cross section; two lowest spin wave guide modes. The phase of $S_{12}$ changes continuously through $\sim 10\pi$ radians for each mode. The coplanar waveguides are coupled by magnetostatic spin waves. For effective separation between the point of excitation and detection $r$, the frequency dependent phase, $\varphi(f)$, can be simply related to a frequency dependent wave vector, $k(f)$, by $\varphi(f) \approx k(f)r + \varphi_p$. Here $\varphi_p$ is an unknown, but assumed constant, phase.

We guide our experimental interpretation with a model calculation of the dispersion, following Arias and Mills[9], for our measured waveguide cross section. We simply adjust $\varphi_p$ for the two modes, indexed as $p$=1, 2, to bring $k(f)$ onto the Arias and Mills dispersion (Figure 2c). The agreement is satisfying and we identify the spin waves with their various modes using the same index $p$.

The magnetostatic spin wave attenuation length is estimated to be $l \sim 3$ μm and determined by comparing results measured on continuous 5 and 10 μm long stripes; it is much shorter than that predicted by eddy currents in this material.[18]

Repeated focused ion beam etching of the same waveguide allows us to introduce and then gradually increase an air gap in the center of the 10 μm long CoTaZr spin wave guide as shown in Figure 1. Figures 3-4 show a ~30% decrease in transmitted amplitude with the first cut followed by a much slower attenuation as the gap is increased from 95 nm to 1.5μm. The p=1 mode of BVMSW is observed at all gap widths while the p=2 mode disappears rapidly.

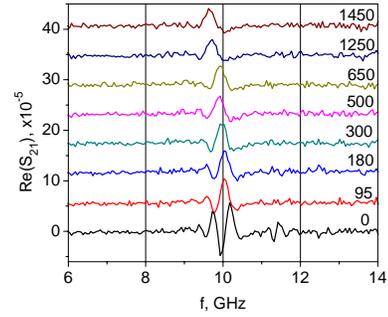

FIG.3. Real part of $S_{21}$ measured on $(10\times1.8\times0.11)$ μm$^3$ CoTaZr spin wave-guide with air gap in the center. Numbers are the gap widths in nanometers.

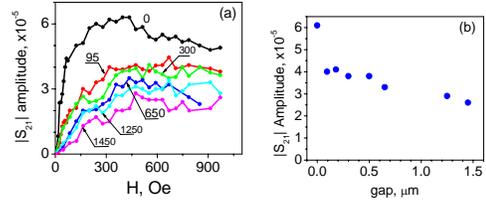

FIG.4. a) Magnetic field dependence of $|S_{21}|$ amplitude measured on $(10\times1.8\times0.11)$ μm$^3$ CoTaZr spin wave-guide with an air gap in the center. Numbers are the gap widths in nanometers. b) Gap width dependence of $|S_{21}|$ peak amplitude measured at H=358 Oe.

We might expect that the dipolar coupling between the two severed ends for the $p$=1 mode will fall dramatically only when the spacing approaches the waveguide width. On the other hand, dipolar coupling of the higher order modes with more rapid phase variation of the oscillating magnetization across the cut face might drop off at gap length scales that exceed the transverse standing wavelengths of the mode - shorter dimension.

Micromagnetic simulations indicate that, despite the shape anisotropy, the magnetization at the ends is not well aligned. As a result, it is not surprising that a modest field is required to turn on the coupling to the co-planar waveguides. Similarly, the static magnetization at the cut ends of the gap will require bias fields to align the end magnetization and turn on the coupling. The data in Figure 4 shows the transmitted signal for gapped waveguides rising more slowly as the gap is widened.

The maximum measured $|S_{12}|$ is displayed as a function of gap width in Figure 4b. The decay of the transmission related to the "tunneling" of the spin waves across the gap is not exponential, as expected, due to the long range nature of dipolar coupling that controls the spin waves[16].

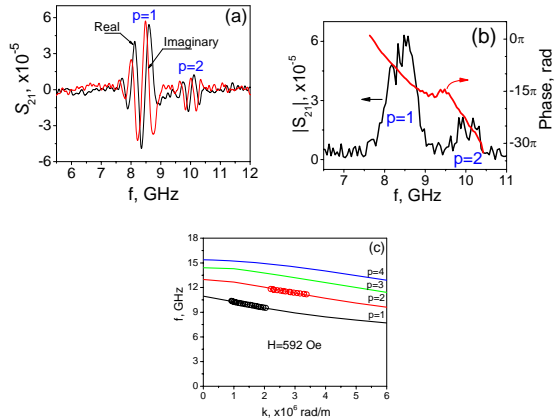

FIG.2. a) Frequency dependence of real and imaginary parts of $S_{21}$; b) Magnitude and phase of $S_{21}$ c) Measured dispersion for $(0.11\times1.8)$μm$^2$ CoTaZr stripes(circles); calculated dispersion for ferromagnetic wires with elliptical profile (solid lines)[9,10]. H=594 Oe.

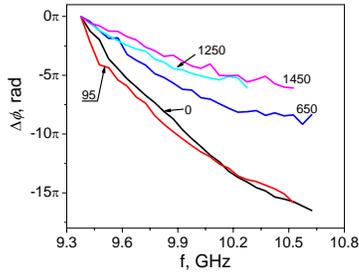

FIG.5. Frequency dependence of the $S_{21}$ phase change measured on $(10\times1.8\times0.11)$ μm$^3$ CoTaZr spin wave-guide with air gap in the center. Numbers represent the gap widths in nanometers.

The $S_{12}$ phase is also affected by the gap; Figure 5 shows the $S_{12}$ phase - frequency dependence change with increase of the gap width. We have no intuition to guide us here and must await models of the long range dipolar mediated gap transmission in order to understand the changes in phase-frequency with gap width.

In summary, we have studied BVMSW modes in micron size CoTaZr wires. Measured spin wave dispersion is in good agreement with model calculations. It was shown that magnetostatic spin wave uniform mode can effectively propagate across the air gap in the ferromagnetic wire. This is not surprising; long range dipolar fields control the magnetostatic wave propagation. Theoretical modeling of magnetostatic spin wave "dipolar tunneling" is required in order to understand the frequency dependent phase shift and transmission amplitude behavior with increase of the gap width.

Magnetostatic spin waves in ferromagnetic metals may be strongly damped by eddy currents that are strongly dependent on material properties such as saturation magnetization, conductivity and geometry[18]. The long range nature of the dipolar forces, which control spin wave propagation, opens the possibility of mitigating eddy current losses by nano-structuring. These results indicate that radical patterning of the ferromagnetic metal wire by introducing nanometer scale air gaps may be a suitable method for suppressing the eddy current losses without disturbing the spin wave phase information.

This work is supported by NERC via the Nanoelectronics Research Initiative (NRI), by Intel Corp. and UC Discovery at the Western Institute of Nanoelectronics (WIN) Center.